\documentclass[superscriptaddress,dvipsnames,nofootinbib,amsmath,amssymb,prd,notitlepage]{revtex4-1}

\usepackage{subfig}
\usepackage[a4paper, left=20mm,right=20mm,top=20mm,bottom=20mm]{geometry}
\usepackage[titletoc,toc,title]{appendix}

\usepackage{mathtools}
\usepackage{tensor}
\usepackage{amsmath, amsfonts,bm}
\usepackage{derivative}
\usepackage{bm}
\usepackage[utf8]{inputenc}
\usepackage[normalem]{ulem}
\usepackage{nicefrac}

\usepackage{suffix}

\usepackage{amsmath, amsfonts,bm}
\usepackage{xcolor}
\usepackage{graphicx}
\definecolor{darkerred}{rgb}{0.9, 0.17, 0.31}
\usepackage{subfig}

\newcommand{\Pre}{\mathcal{P}}

\newcommand{\be}{\begin{equation}}  
\newcommand{\ee}{\end{equation}}
\newcommand{\ba}{\begin{eqnarray}}
\newcommand{\ea}{\end{eqnarray}}

\newcommand{\bem}{\begin{bmatrix}}
\newcommand{\eem}{\end{bmatrix}}
\newcommand{\Mpl}{M_{\rm Pl}}

\newcommand{\mL}{\mathcal{L}}

\newcommand{\mY}{\mathcal{Y}}
\newcommand{\mX}{\mathcal{X}}

\newcommand{\dd}{\text{d}}

\usepackage[colorlinks=true,hyperfootnotes=true, linkcolor=darkerred, citecolor=cyan, urlcolor=magenta]{hyperref}

\DeclareMathAlphabet{\mathup}{OT1}{\familydefault}{m}{n}

\def\dd{\mathrm{d}}

\usepackage[capitalise]{cleveref}

\usepackage{array}
\newcommand{\PreserveBackslash}[1]{\let\temp=\\#1\let\\=\temp}
\newcolumntype{C}[1]{>{\PreserveBackslash\centering}p{#1}}
\newcolumntype{R}[1]{>{\PreserveBackslash\raggedleft}p{#1}}
\newcolumntype{L}[1]{>{\PreserveBackslash\raggedright}p{#1}}
\renewcommand\arraystretch{1.5}

\begin{document}

\title{An obstruction to cosmic acceleration from shift-symmetric dark sector interactions}

\author{Jose Beltrán Jiménez}
\email{jose.beltran@usal.es}
\affiliation{Departamento de F\'isica Fundamental and IUFFyM, Universidad de Salamanca, E-37008 Salamanca, Spain}

\author{Luis J. Garay}
\email{luisj.garay@ucm.es}
\affiliation{Departamento de F\'isica Te\'orica and IPARCOS,
Universidad Complutense de Madrid, 28040 Madrid, Spain}

\author{Erik Jensko}
\email{erik.jensko@ucl.ac.uk}
\email{jenskoerik@gmail.com}
\affiliation{Department of Mathematics, University College London,
Gower Street, London WC1E 6BT, UK}

\author{Mar\'{i}a P\'{e}rez Garrote}
\email{maripe49@usal.es}
\affiliation{Departamento de F\'isica Fundamental and IUFFyM, Universidad de Salamanca, E-37008 Salamanca, Spain}

\begin{abstract}
Using the effective field theory of cosmological media, we show that derivative interactions in the dark sector between a perfect fluid and a superfluid cannot support slow-roll expansion while remaining free from instabilities. Furthermore, decomposing the dark sector medium into effective dark matter and dark energy components, we find that the stability conditions exclude a phantom-to-non-phantom crossing.
Our approach gives a clear route for constructing theoretically viable dark sector models, and highlights the fundamental constraints that stability requirements impose on their cosmological dynamics.
\end{abstract} 

\maketitle

{\hypersetup{linkcolor=black}
\makeatletter
\def\l@subsubsection#1#2{}
\makeatother
\tableofcontents
}

\section{Introduction}
The Lambda-Cold-Dark-Matter model ($\Lambda$CDM) provides a remarkably successful phenomenological model of the Universe, describing observations ranging from the cosmic microwave background (CMB) to the formation of large-scale structure \cite{Planck:2018vyg}. The model relies on two components for describing the dark sector, the cosmological constant $\Lambda$ for the dark energy (DE) component, and CDM for the dark matter (DM) component, and therefore provides a minimal phenomenological framework. However, the fundamental nature of both components remains unknown \cite{Weinberg:1988cp, Sahni:1999gb, Carroll:2000fy, Padmanabhan:2002ji, Peebles:2002gy, Martin:2012bt, Bertone:2004pz}, which leaves open the possibility of a richer dark sector with testable cosmological signatures. Moreover, the long-standing $H_0$ tension \cite{Verde:2019ivm, Riess:2021jrx, DiValentino:2021izs, Freedman:2021ahq} and recent baryon acoustic oscillation (BAO) measurements from the Dark Energy Spectroscopic Instrument (DESI) \cite{DESI:2024mwx, DESI:2025zgx}, provide additional motivation for testing extensions of $\Lambda$CDM \cite{CosmoVerseNetwork:2025alb}.

The DESI DR2 BAO results, when combined with CMB and supernova measurements, favour an evolving equation of state for dark energy which crosses the phantom divide, $w=-1$, from the phantom regime at high redshift to the non-phantom regime at low redshift \cite{DESI:2024mwx, DESI:2025zgx}. For non-interacting dark sector models, crossing the effective phantom divide can present several pathologies \cite{Carroll:2003st, Vikman:2004dc, Hu:2004kh, Kunz:2006wc}. However, modified gravity models \cite{Fujii:1982ms, Wetterich:1987fm, Silvestri:2009hh, DeFelice:2010aj, Clifton:2011jh, Joyce:2014kja, Koyama:2015vza, Ishak:2018his, Kase:2018aps, Koussour:2023ulc, Calderon:2026vgr}, interactions between DM and DE \cite{Pourtsidou:2013nha, Wang:2016lxa, Pourtsidou:2016ico, Kase:2020hst, Asghari:2019qld, Figueruelo:2021elm, BeltranJimenez:2021wbq,Poulin:2022sgp,Jimenez:2024lmm} or dynamical DE models \cite{Chiba:1997ej,Copeland:1997et, Caldwell:1997ii, Hu:2004kh, Copeland:2006wr,Tsujikawa:2013fta, Wolf:2024stt} can circumvent these difficulties and have been explored as alternatives to $\Lambda$CDM. It is therefore crucial to construct theoretically controlled dark sector models capable of explaining current observations while satisfying fundamental stability requirements.

Models beyond the standard perfect-fluid description have been widely investigated in the literature \cite{Hu:1998kj, Arjona:2020yum, Kopp:2016mhm, Malik:2002jb, Malik:2004tf,Celoria:2017bbh, Bemfica:2020zjp}. Among these, superfluids provide an interesting framework for exploring the cosmological implications of media beyond the perfect-fluid paradigm. From the effective perspective, superfluids are particularly appealing because their low-energy dynamics is determined by their internal symmetries: the existence of a conserved $U(1)$ charge together with the spontaneous breaking of the corresponding $U(1)$ symmetry \cite{ISRAEL198179,ISRAEL198277,KHALATNIKOV198270, carter1985canonical, Martin1989,PhysRevD.45.4536,Son:2002zn,Son:2005rv,Escobedo:2010uv,Nicolis:2011pv,Dubovsky:2011sj, Nicolis:2013lma,Nicolis:2023pye}. 
This structure can lead to a richer spectrum of excitations than that of a standard perfect fluid, which only supports a single sound mode. In particular, while at zero temperature the low-energy dynamics is governed by a single propagating superfluid phonon, at finite temperature the interacting medium can be described as a mixture of normal fluid and superfluid components, with different fluid reference frames \cite{Carter:1995if, Son:2005ak, Herzog:2008he, Nicolis:2011cs}. 
These properties have motivated the application of superfluid media to cosmology, such as superfluid dark matter \cite{Berezhiani:2025maf}, where they provide a framework for describing self-gravitating media and its cosmological perturbations \cite{Celoria:2017bbh}.

Relativistic media can be described within the effective field theory (EFT) approach in terms of scalar fields associated with the internal degrees of freedom of the medium, with their dynamics constrained by the underlying symmetries. 
This framework provides a modern counterpart of the well-known Lagrangian fluid actions of Brown \cite{Brown:1992kc} and Schutz-Sorkin \cite{Schutz:1977df}, and encodes the classical thermodynamic properties of the medium in terms of the corresponding scalar invariants \cite{PhysRevLett.94.175301,Endlich:2010hf,Dubovsky:2011sj,Ballesteros:2012kv,Ballesteros:2014sxa,Ballesteros:2016kdx,Celoria:2017hfd}. For a detailed review in EFT for material media, see \cite{Esposito:2026diu}.
In the case of a relativistic perfect fluid interacting with a superfluid component, we require four scalar fields: three fields labelling the comoving elements of the medium, and an additional scalar field associated with the conserved $U(1)$ charge.
The leading order action in the derivative expansion compatible with the internal symmetries can then be constructed in terms of scalar invariants \cite{Dubovsky:2011sj,Celoria:2017bbh}. 
Imposing only internal volume-preserving diffeomorphisms on the three comoving fields gives a barotropic perfect fluid \cite{Ballesteros:2012kv}, while the additional shift-symmetric scalar field introduces the superfluid component.
The EFT is by now well established in cosmology and has been applied to a wide range of relativistic media and cosmological scenarios \cite{Aoki:2025bmj}. Moreover, it provides the kinetic and constraint structure required to identify the propagating degrees of freedom and fully assess their perturbative stability, while remaining agnostic about the
underlying microphysical model.

The EFT description of relativistic media has also been applied to inflationary cosmology, leading to scenarios beyond the standard single-field paradigm, such as solid and supersolid inflation \cite{Endlich:2012pz, Bartolo:2014xfa, Ricciardone:2016lym, Celoria:2020diz, Celoria:2021cxq}, elastic inflation \cite{Gruzinov:2004ty}, and shift-symmetric inflation \cite{Armendariz-Picon:1999hyi, Finelli:2018upr}, as well as inflation driven by adiabatic media \cite{Celoria:2019oiu}. 
In parallel, the EFT of inflation has provided a systematic framework for describing cosmological perturbations around a quasi-de Sitter background \cite{Cheung:2007st}. 
Within these frameworks, stability conditions have been employed to constrain inflationary models and, in some cases, to rule out specific realisations in the slow-roll regime \cite{Creminelli:2006xe,Finelli:2018upr,Celoria:2019oiu,Aoki:2022ylc}.
In this work, we extend this approach to the EFT of a perfect fluid interacting with a superfluid, applying the stability conditions during slow-roll evolution and establishing general no-go results for this class of imperfect media. In particular, we show that  slow-roll solutions are at odds with stability conditions. Since a prolonged quasi-de Sitter phase with $w\simeq-1$ is relevant for both inflation and late-time cosmic acceleration, our results have  implications for both early- and late-time cosmology. However, while slow-roll is excluded, a stable approach to de Sitter remains possible.

By decomposing the medium into an effective dark matter and an effective dark energy component, one can naturally regard our formalism as describing an interacting dark sector, 
with non-trivial derivative couplings between the two components at the linear perturbation level. Given the relevance of phantom crossing to recent cosmological observations, we explore whether this phenomenon can be realised by our interacting two-component medium. This leads to additional no-go results: the total medium cannot allow for stable phantom-divide crossing, extending known results for barotropic fluids \cite{Scherrer:2005je,Linder:2008ya}, and an effective dark energy component can only cross from the non-phantom to the phantom regime. The latter result is especially important in light of the preference of DESI-based analyses for an effective dark energy component which evolves from the phantom regime at higher redshift to the non-phantom regime at low redshifts \cite{DESI:2024mwx, DESI:2025zgx}. 
Our restrictions show that a stable crossing may only evolve in the opposite direction.
Lastly, we explore the restrictions that our stability conditions impose on existing dark sector models using the EFT formalism already present in the literature \cite{Celoria:2017xos}.

This paper is organised as follows. In Sec.~\ref{sec:eft}, we introduce the effective Lagrangian for a perfect fluid   interacting with a superfluid and derive the classical thermodynamic quantities in a covariant manner. In Sec.~\ref{sec:cosmology}, we study the dynamics of the medium in an FLRW background and analyse its scalar perturbations to obtain the corresponding stability conditions. In Sec.~\ref{sec:nogo}, we apply these conditions to slow-roll expansion and phantom crossing. We also discuss the theoretical viability of several dark sector models. Finally, we present our conclusions in Sec.~\ref{sec:conclusions}.

\section{Effective fluids in cosmology} \label{sec:eft}
We begin by introducing the theoretical framework used to describe the interacting medium in cosmology within the EFT language. This framework allows us to formulate the Lagrangian from operators built from the low-energy cosmological degrees of
freedom, which encode quantities such as number density, entropy, and superfluid properties, with symmetry used
as a guiding principle. The Lagrangian will provide the kinetic and constraint structure
required to identify the propagating degrees of freedom and assess the stability of perturbations.
After introducing the necessary concepts to formulate the perfect fluid interacting with the superfluid in the EFT approach, we derive the fully covariant dynamics and effective fluid quantities. We provide a further illustrative mapping between the EFT variables and thermodynamical quantities in Appendix \ref{Mapping}, giving a more fundamental interpretation of the EFT quantities. See also the recent comprehensive review \cite{Esposito:2026diu}.

\subsection{Theoretical framework}

In this section, we review the theoretical formulation of relativistic superfluids and perfect fluids. To this end, we employ the language of effective field theory (EFT) \cite{Nicolis:2011cs, Celoria:2017bbh}. A general fluid can be described in terms of four scalar fields $\phi^0$ and $\phi^a$ (with $a=1,2,3$). In this approach, we assign three labels $\phi^a$ to each individual volume element of the fluid. In a spacetime coordinate system $x^\mu=(t,\vec{x})$, the trajectory of such a fluid element in physical space is given by $\vec{x}(\phi^a,t)$. The inverse map $\phi^a(t,\vec{x})$ describes how fluid elements propagate through space, so that $\phi^a$ can be interpreted as the comoving coordinates of the fluid \cite{Dubovsky:2005xd, Ballesteros:2012kv, Ballesteros:2016kdx}. 
The fourth scalar $\phi^0$ is essential for including features beyond those of a perfect fluid, see for instance \cite{Son:2005ak,Ballesteros:2016kdx,Celoria:2017bbh, Celoria:2017xos}. It can be interpreted as an internal time coordinate of the medium or the field that carries a conserved charge \cite{Dubovsky:2011sj}.

A defining property of a fluid is that its volume elements can be rearranged internally without changing the energy, provided that the total volume remains unchanged. In the EFT language, this corresponds to imposing invariance under volume-preserving internal diffeomorphisms
\be
\phi^a\rightarrow \zeta^a(\phi),\qquad \text{det}\left(\frac{\partial \zeta^a}{\partial \phi^b} \right)=1\,.
\label{volume diff}
\ee
This symmetry corresponds to an internal Transverse Diffeomorphisms invariance (TDiff), which contains the ISO(3) subgroup that describes the homogeneity and isotropy of cosmological scenarios. Apart from the volume-preserving internal diffeomorphisms invariance for the fields $\phi^a$, we require the fourth scalar $\phi^0$ to satisfy a shift symmetry
\be
\phi^0\to\phi^0+c \, ,
\label{shifts}
\ee
where $c$ is a constant parameter. We can define a four-velocity of the medium $u^\mu$ through the conditions $u^\mu\partial_\mu\phi^a=0$ and $u^\mu u^\nu g_{\mu\nu}=-1$,
whose only future-directed solution is \cite{Dubovsky:2005xd, Endlich:2010hf, Endlich:2012pz}
\be\label{four vel}
u^\mu=\frac{1}{6\sqrt{b}\sqrt{-g}}\epsilon^{\mu\alpha\beta\gamma}\epsilon_{abc}\partial_\alpha\phi^a\partial_\beta\phi^b\partial_\gamma\phi^c \, ,
\ee
where $\epsilon^{\mu\alpha\beta\nu}$ and $\epsilon_{abc}$ are the four- and three-dimensional Levi-Civita symbols, with $\epsilon^{0123}=+1=\epsilon^{123}$. The projector on hypersurfaces orthogonal to the four-velocity has the form
\be
h_{\mu\nu}=g_{\mu\nu}+u_\mu u_\nu,\qquad h_{\mu\nu}u^\nu=0\,.
\label{projector}
\ee

Imposing invariance under volume-preserving internal
diffeomorphisms \eqref{volume diff} and shift symmetry \eqref{shifts} at leading order in the derivative expansion, the Lagrangian can only depend on three independent scalars
\cite{Nicolis:2011cs}
\be
\mL= -F(b,\mY,\mX)\,,
\label{fluLag}
\ee
with
\be
b:=\text{det}B^{ab},\qquad \mY:= u^\mu \partial_\mu\phi^0\,, \qquad \mX:= g^{\mu\nu}\partial_\mu\phi^0\partial_\nu\phi^0\,,
\label{scalars}
\ee
and where
\be
B^{ab}= g^{\mu\nu}\partial_\mu\phi^a\partial_\nu\phi^b \,, 
\ee
is the induced metric of the medium. We can also define another independent four velocity associated with the irrotational component of the superfluid \cite{Ballesteros:2016gwc}
\be
\omega^\mu=-\frac{\partial^\mu \phi^0}{\sqrt{-\mX}}.
\label{velw}
\ee

From the Lagrangian \eqref{fluLag} we can identify some relevant subcases. The determinant $b$ measures the local change in the volume of the medium and is therefore associated with the fluid's density degree of freedom. As we will see in more detail in the next section, the case in which $F$ depends only on $b$ corresponds to a barotropic perfect fluid. On the other hand, $\mX$ is the kinetic term of the scalar field $\phi^0$, so a Lagrangian which depends only on $\mX$ describes the purely kinetic \textit{k-essence} theories that provide the EFT of a superfluid \cite{Ballesteros:2016kdx}, which is nothing but an irrotational barotropic perfect fluid \cite{Dubovsky:2005xd, Arroja:2010wy}. The invariant $\mY$ couples the fluid four-velocity $u^\mu$ to the gradient of the scalar field $\partial_\mu\phi^0$ so it can be interpreted as an entrainment coupling \cite{Andersson:2006nr,Ballesteros:2013nwa}. In this setup, $b$ therefore describes the barotropic perfect-fluid sector, $\phi^0$ the superfluid, and $\mY$ their entrainment coupling, so that $F(b,\mY,\mX)$ describes a perfect fluid interacting with a superfluid component.  In fact, we will see explicitly that Lagrangians of the form $F(b)$, $F(\mY)$ and $F(\mX)$ all describe barotropic fluids. The perfect fluid described by a Lagrangian that does not depend on $\mX$, i.e. of the form $F(b, \mY)$, is in general non-barotropic, since it depends on two independent scalars, and can have non-adiabatic pressure perturbations. Note that in this case, although the entropy per particle is conserved along the fluid flow, it can still differ between fluid elements. This allows for spatially-varying intrinsic entropy perturbations even within a perfect fluid \cite{Celoria:2017xos}. Some other special cases of the general superfluid are discussed in \cite{Ballesteros:2016gwc}. In the next section, we describe the properties of these distinct subcases in more detail.

For the gravitational sector, we adopt the standard Einstein-Hilbert Lagrangian so that the total action is
\be
{\cal S}=\int {\rm d}^4 x \sqrt{-g}
\left[ \frac{R}{16 \pi G}-F(b,\mY,\mX) \right]\,,
\label{action}
\ee
with $\Mpl^{-2}=8\pi G$.
The action \eqref{action} describes a perfect fluid interacting with a superfluid component minimally coupled to gravity within the EFT framework, where the degrees of freedom are parametrized by four scalar fields obeying the aforementioned internal symmetries. All the physical quantities can be obtained from the Lagrangian \eqref{fluLag}, e.g., the pressure and energy density, the conserved currents, the background equations of motion and the properties of the perturbations.

\subsection{Covariant dynamics} \label{sec:mapping}

The energy-momentum tensor associated with the Lagrangian \eqref{fluLag} is
\be
T_{\mu\nu}:=-2\frac{\partial \mL}{\partial g^{\mu\nu}}+\mL g_{\mu\nu}=2\left(F_b \frac{\delta b}{\delta g^{\mu\nu}} + F_\mY \frac{\delta \mY}{\delta g^{\mu\nu}}+ F_\mX \frac{\delta \mX}{\delta g^{\mu\nu}}\right)-Fg_{\mu\nu}\,,
\label{Tmunu}
\ee
where $F_b$, $F_\mY$ and $F_\mX$ are the derivatives of $F(b,\mY,\mX)$ with respect to $b$, $\mY$ and $\mX$, respectively. Using Eqs.~(\ref{scalars}) and (\ref{velw}) we can rewrite the energy-momentum tensor as
\be
T_{\mu\nu}
=\left(F-\mY F_\mY\right)u_\mu u_\nu-2\mX F_\mX \omega_\mu \omega_\nu+\left(2bF_b-F\right)h_{\mu\nu}\,,
\label{Tmunu2}
\ee
where we have used Eq.~(\ref{projector}) and $B_{ab}^{-1}\partial_\mu \phi^a\partial_\nu \phi^b=h_{\mu\nu}$ for the inverse of $B^{ab}$ \cite{Dubovsky:2011sj}. 

If the Lagrangian does not depend on the scalar $\mX$, the energy-momentum tensor reduces to the usual perfect fluid case
\be
T_{\mu\nu}
=\left(F-\mY F_\mY\right)u_\mu u_\nu+\left(2bF_b-F\right)h_{\mu\nu}\,,
\ee
with only one reference frame along $u^\mu$. As discussed in Ref.~\cite{Nicolis:2011cs}, Eq.~(\ref{Tmunu2}) contains two components in relative motion with four-velocities $u^\mu$ and $\omega^\mu$. In particular, the relative velocity of the superfluid component, \cref{velw}, with respect to the fluid, \cref{four vel}, is given by the vector \cite{landau1987fluid, Nicolis:2011cs, Celoria:2017bbh}
\be
\xi_\mu=-\mY u_\mu + \sqrt{-\mX}\omega_\mu\,,
\label{xi}
\ee
with $\xi\equiv\sqrt{\xi_\mu \xi^\mu}=\sqrt{\mY^2+\mX}$.
We will see in the next section that the modulus of the relative velocity vanishes in the particular FLRW background where the cosmological principle stipulates a common rest frame for all the components on large scales. When there is only a single fluid component moving along a fluid frame, as in the case of the perfect fluid $F(b,\mY)$ and the barotropic fluids $F(\mY)$ and $F(\mX)$, $\xi$ is simply a renormalisation of the fluid velocity. To define the thermodynamic quantities, we choose the reference frame along the fluid flow $u^\mu$ orthogonal to the internal space $u^\mu\partial_\mu\phi^a=0$. In a cosmological context, this will be important for the perturbations, but  this choice is not crucial for the background evolution since, as explained, all components share the same rest frame.

The energy-momentum tensor of a generic medium with respect to $u^\mu$ is given by \cite{landau1987fluid, Ellis:1998ct}
\be
T_{\mu\nu}=\rho u_\mu u_\nu+q_\mu u_\nu+q_\nu u_\mu+\mathcal{P} h_{\mu\nu}+\pi_{\mu\nu}\,,
\label{T medium}
\ee
where $\rho$ is the energy density, $\mathcal{P}$ is the pressure, $\pi_{\mu\nu}$ is the trace-free anisotropic pressure and $q_\mu$ is the energy flux. We identify the energy density and the pressure of the energy-momentum tensor from \cite{Ellis:1998ct}
\be
\rho=T_{\mu\nu}u^\mu u^\nu=F-\mY F_\mY+2\mY^2 F_\mX\,,\qquad
\Pre=\frac{1}{3}h^{\mu\nu}T_{\mu\nu}=2bF_b-F+\frac{2}{3}\left(\mY^2+\mX\right)F_\mX\,,
\label{rhoandP}
\ee
where we have used Eqs.~(\ref{fluLag}) and (\ref{Tmunu2}). 
The anisotropic pressure in the fluid frame is obtained from the orthogonally projected, symmetric, trace-free part of the energy-momentum tensor \eqref{T medium} \cite{Ellis:1998ct}
\be
\pi_{\mu\nu}=T_{\langle\mu\nu\rangle}=\left[-h_{\mu\nu}\frac{2}{3}\left(\mY^2+\mX\right)+2\xi_\mu\xi_\nu\right]F_\mX\,,
\label{anisotropic}
\ee
which satisfies $\pi^{\mu}{}_{\mu}=0$ and $\pi_{\mu\nu}u^\nu=0$. This expression is proportional to $F_\mX$, thus vanishes when the Lagrangian does not depend on $\mX$. It is clear from \cref{anisotropic} that the relative-motion stress is \mbox{$\pi_{\mu\nu}+\nu h_{\mu\nu}=2F_\mX\xi_\mu\xi_\nu$}, so its trace part, which is contained in $\Pre$ in \cref{rhoandP}, is the viscosity term $\nu=\frac{2}{3}\left(\mY^2+\mX\right)F_\mX$. Therefore, we have that the thermodynamic pressure in the fluid frame is
\be
P=\Pre-\nu=2bF_b-F.
\label{isoP}
\ee

With this identification we can see that the energy-momentum tensor \eqref{Tmunu2} is the sum of an isotropic term $Ph_{\mu\nu}$ and two terms corresponding to the two fluid components $u^\mu u^\nu$ and $\omega^\mu \omega^\nu$.
Finally, the energy flux relative to $u^\mu$ is
\be
q^\mu=-h^{\mu\alpha}u^\beta T_{\alpha\beta}
=2h^{\mu\alpha}\omega_\alpha\sqrt{-\mX}\mY F_\mX\,,
\label{heatflux}
\ee
which satisfies $q_\mu u^\mu=0$. We can see from Eqs.~(\ref{rhoandP}),  (\ref{isoP}) and (\ref{heatflux}) that the superfluid component $\mX$ not only contributes to the energy density, the pressure and the anisotropic pressure, but is also the source of the energy flux in the fluid frame.

If the Lagrangian does not depend on $\mX$, the energy flux and the anisotropic stress vanish and the energy-momentum tensor \eqref{Tmunu2} reduces to the non-barotropic perfect fluid tensor \mbox{$T_{\mu\nu}=\rho u_\mu u_\nu + P h_{\mu\nu}$}, with $\rho$ and $P$ depending on two variables $b$ and $\mY$. We recover the barotropic perfect fluid as a special case when the Lagrangian depends only on one of these two variables, i.e. when the Lagrangian is of the form $F(b)$ or $F(\mY)$. The Lagrangian $F(\mX)$ also describes an irrotational and barotropic perfect fluid in the frame along $\omega^\mu$ with \mbox{$T_{\mu\nu}=-2\mX F_\mX \omega_\mu \omega_\nu-F g_{\mu\nu}$}.

There are two different types of conserved currents present in the medium we are considering. The current
\be
J^\mu_n=\sqrt{b}u^\mu\,,
\label{density current}
\ee
is conserved off-shell, i.e. $\nabla_\mu J^\mu_n=0$ independently of any symmetry choice of the Lagrangian. Since the four-velocity $u^\mu$ is related to the internal volume element, see Eq.~(\ref{four vel}), the current \eqref{density current} is interpreted as the current of fluid points \cite{Dubovsky:2005xd}. We therefore define the number density as
\be
n:= -J_n^\mu u_\mu=\sqrt{b}\,.
\label{n}
\ee
The off-shell conserved current \eqref{density current}
is then interpreted as the conserved number density current in this thermodynamical representation.
On the other hand, there are currents coming from the symmetries we have imposed in Eq.~(\ref{fluLag}). The conserved Noether current corresponding to the shift symmetry \eqref{shifts} is
\be
J^\mu_s=\frac{\partial \mL}{\partial \left(\partial_\mu \phi^0 \right)}=-F_\mY u^\mu-2F_\mX\partial^\mu \phi^0\,,
\label{entropy current}
\ee
and it is conserved on-shell. We can then define the entropy density
\be
s:=-J_s^\mu u_\mu=2\mY F_\mX-F_\mY\,.
\label{s}
\ee
Using Eq.~(\ref{heatflux}), the Noether current can be written as \cite{Celoria:2017bbh}
\be
J_s^\mu = s u^\mu + \frac{q^\mu}{\mY}\,,
\ee
which is the entropy current of an imperfect fluid \cite{Rezzolla:2013dea} conserved on-shell.  

Finally, in Appendix \ref{Mapping}, we review the dictionary connecting classical thermodynamical quantities and EFT of superfluids, following Refs. \cite{Celoria:2017bbh, Ballesteros:2016kdx}. This lets us naturally identify the operator $\mY$ as the temperature of the medium
$T \equiv \mY$.
Such an identification will be particularly useful for giving a thermodynamical interpretation of the stability conditions in the next section.

\section{Cosmology} 
\label{sec:cosmology}
Having reviewed the thermodynamic interpretation of the interacting superfluid medium and derived the most relevant physical quantities in a covariant manner, we now turn to the study of its cosmological applications. We first study the background cosmological equations before examining scalar perturbations and, finally, deriving the corresponding stability conditions and propagation speeds in the small scale limit. For simplicity, we do not consider the contribution from the standard matter components, since including them does not modify the stability conditions.

\subsection{Background equations}
Let us study the dynamics of the coupled perfect fluid and superfluid on a spatially flat Friedmann-Lema\^itre-Robertson-Walker (FLRW) background parametrised by the following line element:
\be
\text{d}s^2=a(\eta)^2\left(-\text{d}\eta^2 + \delta_{ij}\text{d}x^i \text{d}x^j\right),
\ee
where $a$ is the scale factor and $\eta$ is conformal time. The background fluid configuration compatible with homogeneity and isotropy of the FLRW metric can be parametrised as
\be
\phi^a= \lambda x^a\,,\qquad\phi^0=\chi\,,
\label{back_coordinates}
\ee
where $\lambda$ is a constant parameter and $\chi=\chi\left(\eta\right)$ is an arbitrary function of conformal time. The homogeneity and isotropy of $\phi^0$ is trivial, while the configuration for $\phi^a$ breaks both homogeneity and isotropy. However, the internal volume preserving diffeomorphism invariance permits to realise a diagonal Euclidean invariance where spatial rotations and translations are compensated by an internal  TDiff transformation. This is possible because the Euclidean group is a subgroup of TDiffs. On this background configuration the EFT scalars become
\be
b=\frac{\lambda^6}{a^6}\,,\qquad \mY=\frac{\chi'}{a}\,,\qquad \mX=-\frac{\chi'^2}{a^2}\,,
\label{back scalars}
\ee
where $'$ denotes derivative with respect to conformal time. Since the parameter $\lambda$ can be absorbed by rescaling the spatial coordinates, we set $\lambda=1$ without loss of generality\footnote{We do not include additional matter sectors in this work, as minimally coupled species do not modify the stability conditions derived below. The presence of baryons, for instance, may affect the normalization of peculiar velocities, in which case the parameter $\lambda$ must be retained.}. From \eqref{back scalars} we have $\mX=-\mY^2$, therefore $\mY$ and $\mX$ are not independent variables at the background level. This relation guarantees that $\xi=\sqrt{\mY^2+\mX}=0$, and additionally $\omega^\mu=u^\mu=\left(a^{-1},0,0,0\right)$ for $\chi'>0$ (i.e. positive temperature $T=\mY>0$).
From now on, the function $F$ and all its derivatives are understood to be evaluated on the background configuration \eqref{back scalars} to alleviate the notation. 

The background energy density and pressure are
\be
\rho=F-\frac{\chi'}{a}F_\mY+2\frac{\chi'^2}{a^2}F_\mX\,,\qquad
P=\frac{2}{a^6}F_b-F\,,
\label{rhoandP_back}
\ee
and the background equation of state is
\be
w:=\frac{P}{\rho}=\frac{-1+2\frac{bF_b}{F}}{1-\frac{\mY F_\mY}{F}-2\frac{\mX F_\mX}{F}}.
\label{eq_of_state}
\ee
As expected,
when $F$ is dominated by a term that depends only weakly on $b$, $\mY$ and $\mX$ the background approaches a quasi-de Sitter phase with $w\simeq-1$. 

Let us now study the background dynamics. The fluid sector characterised by the variable $b$ does not have an associated equation of motion since the configuration $\phi^a=x^a$ corresponds to a trivial solution. The dynamics of this sector is encoded in the functional dependence of $F$ on $b$. Thus, only $\phi^0$ will have a dynamical evolution governed by the corresponding equation of motion that can be obtained by varying the action \cref{action} with respect to $\chi$ on the background. The resulting equation reads
\be\label{eq chi}
\chi''=\mathcal{H}\chi'\left(1-3\beta\right)\,,
\ee
where $\mathcal{H}=a'/a$ is the Hubble expansion rate in conformal time and we have defined the function
\be
\beta=\frac{F_\mY-2bF_{b\mY}+4b\mY F_{b\mX}- 2\mY F_\mX}{\mY \left(F_{\mY\mY}-4\mY F_{\mY\mX}-4\mX F_{\mX\mX}-2F_\mX\right)},
\ee
that captures the deviation from the standard Hubble friction term. 
Equivalently, the equation \eqref{eq chi} can be expressed as the conservation of the Noether current associated with the $\phi^0$ shift symmetry, i.e., the entropy density current conservation $\nabla_\mu J^\mu_s=0$ with $J^{\mu}_s$ in \cref{entropy current}. On FLRW backgrounds this takes the compact form
\be\label{entropy_cons}
\frac{1}{a^3}\frac{\dd}{\dd\eta}
\left[
a^3\left(2\mY F_\mX-F_\mY\right)
\right]
=0 \, .
\ee
Solving \cref{entropy_cons} for $\chi''$ gives precisely \cref{eq chi}. This agrees with the background entropy per particle \mbox{$\sigma=2a^2\chi'F_\mX-a^3F_\mY$} being conserved along the fluid flow $u^\mu \nabla_\mu \sigma=0$, since the energy flux \eqref{heatflux} vanishes on the FLRW background, see Appendix \ref{Mapping}.

The Friedmann equations governing the background evolution are
\ba \label{Fried1}
    \mathcal{H}^2&=&\frac{a^2}{3\Mpl^2}\left(F-\mY F_\mY-2\mX F_\mX\right)\,,\label{eqsback} \\
    \mathcal{H}'&=&\frac{a^2}{6\Mpl^2}\left(2F+\mY F_\mY-6bF_b+2\mX F_\mX\right)\,. \label{eqsback2}
\ea
These equations imply the background continuity equation
\be \label{cont_eq}
    \rho' + 3\mathcal{H} (\rho+P) =0 \, ,
\ee
which holds on-shell. 
Note that this is an off-shell identity only in the barotropic case $F(b)$, following from the choice of configuration $\phi^{a}=x^a$, while in the general case it follows from the Noether current enforcing entropy conservation \cref{entropy_cons}.

\subsection{Scalar perturbations and stability} \label{sec:cosmo:stab}

After introducing the background configuration of our cosmological scenario, we will turn our attention to the scalar perturbations in order to obtain the stability conditions associated with the absence of ghosts and Laplacian
instabilities. In this section, we present the quadratic-order action obtained by expanding the action up to second order in scalar perturbations, followed by the stability analysis. We analyse the vector and tensor perturbations in the Appendix \ref{Append1}. We focus on the regime dominated by the interacting perfect fluid and superfluid and neglect the contribution from the rest of the standard components. Note that including minimally coupled matter components does not modify the stability conditions, since the perfect fluid–superfluid system couples to the remaining components through gravity. The stability conditions obtained in this section agree with those presented in \cite{Celoria:2017hfd}, and our leading-order propagation speeds match the flat-space results originally derived in 
\cite{Nicolis:2011cs}. In the next section, we utilise these constraints to derive our main results, namely: no slow-roll cosmological scenarios are compatible with the absence of instabilities. We will also prove that an effective crossing of the phantom divide line can only occur in one direction.

In order to obtain the evolution of scalar linear perturbations, we will parametrise the perturbations around the configuration \eqref{back_coordinates} as
\be
\phi^a=x^a+\pi^a\,,\qquad\phi^0=\chi+\pi_0\,,
\ee
where $\pi^a$ and $\pi_0$ describe the phonon perturbations of the perfect fluid and superfluid components. We can split the space perturbation $\pi^a$ into transverse and longitudinal parts
\be
\pi^a=\pi^a_T+\partial^a\pi_L,
\ee
with $\partial_a\pi^a_T=0$. The transverse component $\pi_T$ corresponds to vector perturbations under the unbroken SO(3) group of the background, while $\pi_L$ and $\pi_0$ belong to the scalar sector. At this order and given the background configuration, we can identify internal and spatial indices and we will use the Euclidean metric to raise and lower internal indices, e.g., $\partial_a \pi_L=\delta_{ab}\partial^b \pi_L$.

For the metric, we consider the perturbed line element \cite{MUKHANOV1992203}
\be
{\rm d}s^2 = -a^2(1 + 2A){\rm d}\eta^2 
+ 4a^2 \partial_{i}B{\rm d}\eta {\rm d}x^i
+ a^2 \Big[ (1 - 2\zeta) \delta_{ij}
+ 2 \left(\partial_i \partial_j -\frac13 \delta_{ij}\partial^2\right) E  \Big]
{\rm d}x^i {\rm d}x^j \,,
\label{pertmetric}
\ee
where $A$, $B$, $\zeta$, and $E$ are scalar perturbations.
Expanding the total action \eqref{action} up to quadratic order, integrating out the boundary terms, and using the background equations of motion, the resulting quadratic action for scalar perturbations reads
\be
{\cal S}^{(2)}_{\text{scalar}}=\int {\rm d}^4 x 
\left( {\cal L}_{\rm flat}+{\cal L}_{\zeta}
+{\cal L}_{E} \right)\,,
\label{pert action}
\ee
where 
\ba
{\cal L}_{\rm flat} &=& a^2 C_1 \left(\pi_0'\right)^2-a^2F_\mX\left(\partial\pi_0\right)^2
+a^4\left(bF_b-\frac12 \mY F_\mY \right)\left(\partial\pi_L'\right)^2-a^4bP_b\left(\partial^2\pi_L\right)^2
\nonumber \\
& &
+a^3F_\mY\partial\pi_L'\partial\pi_0 + 2a^3bs_b\partial^2\pi_L\pi_0'
+4a^4\left(bF_b-\frac12 \mY F_\mY \right)\left(\partial B \right)^2-a^4\left(\rho-\mY^2 C_1\right)A^2
\nonumber \\
& &
-2a^4b \rho_{b} \partial^2\pi_L A
-2\mY a^3 C_1\pi_0' A
-4\mY a^3F_\mX \partial \pi_0\partial B
\nonumber \\
& &
-4 a^4\left(bF_b-\frac12 \mY F_\mY \right)\partial\pi_L'\partial B
+\Mpl^2a^2\left(\partial^2 B A'-\partial B'\partial A-2\mathcal{H}\partial^2B A
\right)\, ,
\label{L_flat}
\\
{\cal L}_{\zeta} &=&-3\Mpl^2a^2\left(\zeta'\right)^2 + \Mpl^2a^2\left(\partial\zeta\right)^2 -3 a^4 C_2\zeta^2 -\Mpl^2 a^2  \left[6\mathcal{H}\left(A+\zeta\right)+\partial^2 B \right]\zeta'
\nonumber \\
& &
+3\Big[a^3 (2 b s_b -s )\pi_0' -2a^4 b \rho_b A
-2a^4 bP_b\partial^2\pi_L-\frac23 \Mpl^2a^2\mathcal{H}\partial^2 B
\Big]\zeta
-\Mpl^2 a^2 \left(8\mathcal{H}\partial B +3\partial B'+2\partial A \right)\partial\zeta\,,
\nonumber
\\
{\cal L}_{E} &=& \frac13 \Mpl^2 a^2 \left(\partial^2 E'\right)^2+ \frac{2}{3} a^4\left(\rho-P \right)\left(\partial^2 E\right)^2+\frac19 \Mpl^2 a^2 \partial(\partial^2 E)\partial(\partial^2E)
\\
& &
+\frac23 \Mpl^2 a^2 \left(4\mathcal{H}\partial^2 B + \partial^2 A - \partial^2 \zeta \right)\partial^2 E + \frac43 \Mpl^2 a^2 \left(2\mathcal{H}\partial^2 E - \partial^2 B \right)\partial^2 E' - \frac83 \Mpl^2 a^2 \mathcal{H} \partial^2 B \partial^2 E \,,
\ea
and we use the shorthand notation $\partial^2 A=\partial_i\partial^i A$ and $\partial A \partial B= \partial_i A\partial^i B$.
We have also used the thermodynamic derivatives 
\begin{equation}
    \rho_{b}:= \frac{\partial \rho}{\partial b} = F_b - \mY F_{b \mY}+ 2 \mY^2 F_{b\mX} \,, \qquad P_{b} := \frac{\partial P}{\partial b}=F_b+2bF_{bb}\,, \qquad 
    s_b:= \frac{\partial s}{\partial b} = 2\mY F_{b\mX} -F_{b\mY}\,,
\label{thermo_ders}
\end{equation}
and defined the coefficients 
\begin{equation} \label{eq:coefs}
     C_{1} := F_\mX- \frac{1}{2} F_{\mY\mY}+2\mY F_{\mY\mX}-2\mY^2F_{\mX\mX}\, , \qquad C_{2} := F+2bF_b+\mY^2 F_\mX-\frac12 \mY F_\mY+6b^2 F_{bb}\,.
\end{equation}
This expression generalises the quadratic action computed in flat space-time \cite{Nicolis:2011cs, Ballesteros:2012kv, Celoria:2017hfd, Comelli:2022evf}.
Interestingly, the coefficient $C_1$ also has a natural thermodynamic interpretation. Using the thermodynamic definitions in \cref{Tabla1} of Appendix \ref{Mapping}, we can define the free energy density  $f(n,T,\xi):=F\left(n^2,T,\xi^2-T^2 \right)$ where $b=n^2$, $\mY=T$ and $\mX=\xi^2-T^2$. Since $s=-(\partial f/\partial T)=2\mY F_\mX-F_\mY$,
the heat capacity per unit volume is \cite{callen1960thermodynamics}
\begin{align}
\label{eq:heatcap}
C_{V} :=
T\left(\frac{\partial s}{\partial T}\right)_{n,\xi}
=-T\left(\frac{\partial^2 f}{\partial T^2}\right)_{n,\xi} = T\left(\partial_\mY-2\mY\partial_\mX\right)
  \left(2\mY F_\mX-F_\mY\right)
=2TC_1 \,,
\end{align}
with derivatives taken at fixed
particle density $n$ and relative velocity $\xi$. 
This definition is also consistent with $\rho=f+Ts$ and the first law of thermodynamics in \cref{diffthermo}, see \cite{Ballesteros:2016kdx} for an analogous discussion in the perfect fluid limit. For $T>0$, the condition $C_1>0$ is  equivalent
to a positive heat capacity $C_V>0$.

The equations of motion for $A$, $B$, $\zeta$, $E$, $\pi_L$ and $\pi_0$ can be obtained by varying \eqref{pert action} with respect to these fields. From the quadratic action, we can directly obtain the stability conditions associated with the absence of ghosts and Laplacian instabilities. We fix the gauge to the Spatially Flat Slicing Gauge (SFSG) defined by $\zeta=E=0$ \cite{Kodama:1984ziu}. Using the perturbation equations of motion we can integrate out the non-dynamical fields $A$ and $B$ from the second-order action \eqref{pert action}, so the scalar reduced action only contains the dynamical perturbations $\{\pi_L,\pi_0\}$. Going to Fourier space, see Appendix \ref{Matrix components}, the quadratic action for the dynamical degrees of freedom can be written in the form \cite{Celoria:2017hfd, Aoki:2025bmj} 
\be\label{eq:quadratic_action}
{\cal S}^{(2)}_{\text{scalar}}=\int \frac{{\rm d} \eta{\rm d}^3 k}{(2\pi)^3}a^4\left( 
\vec{\mathcal{X}}^{t}_{\mathbf{k}}{}'{\bm K}\vec{\mathcal{X}}{}'_{-\mathbf{k}}
-\frac{k^2}{a^2}\vec{\mathcal{X}}^{t}_{\mathbf{k}}{\bm G}\vec{\mathcal{X}}_{-\mathbf{k}}
-\frac{k}{a}\vec{\mathcal{X}}^{t}_{\mathbf{k}}{\bm B}\vec{\mathcal{X}}{}'_{-\mathbf{k}}
\right)\,,
\ee
where ${\bm K}$ and ${\bm G}$ are $2 \times 2$ symmetric matrices and we have performed integration by parts to define ${\bm B}$ as an antisymmetric $2 \times 2$ matrix with derivative-mixing cross terms. The two dynamical perturbations $\{\pi_L,\pi_0\}$ are represented by
\be
\vec{\mathcal{X}}^{t}_{\mathbf{k}}=\left(k\pi_L(\mathbf{k}), 
\pi_0 (\mathbf{k})\right) \,.
\ee
We have normalized the field $\pi_L$ to have the canonical order $\mathcal{O}(k^0)$ terms in the diagonal components of the matrix ${\bm K}$. The explicit form of the matrix components are given in Appendix \ref{Matrix components}.

The absence of ghost-like instabilities requires ${\bm K}$ to be positive definite, which is guaranteed by imposing $K_{11} > 0$ 
and ${\rm det}\,{\bm K} = K_{11}K_{22} - K_{12}^2 > 0$. Although these conditions should be satisfied at all scales, we will focus on the UV regime where simpler conditions can be obtained. Thus, our conditions will be necessary, but they may not be sufficient. Since $K_{12}$ is suppressed relative to $K_{11}$ and $K_{22}$ in the large-$k$ limit, the ghost-free conditions reduce to $K_{11} > 0$ and $K_{22} > 0$, namely
\be \label{eq:no_ghost}
bF_b-\frac12 \mY F_\mY>0\,, \qquad  C_1 = F_\mX -\frac12 F_{\mY\mY} + 2\mY F_{\mY\mX} - 2\mY^2F_{\mX\mX} >0 \,.
\ee
Using the definition of the heat capacity $C_V=2TC_1$ from \cref{eq:heatcap}, we obtain a direct thermodynamic interpretation of the second no-ghost condition: for positive temperature superfluids $T>0$, we require a positive heat capacity $C_{V}>0$. This extends the perfect fluid connections between thermal and dynamical stability discussed in \cite{Ballesteros:2016kdx} to the general superfluid case. 

After deriving the no-ghost conditions, we turn our attention to the presence of Laplacian instabilities. For the same reasons as explained above, we will obtain necessary conditions from the UV regime. Taking the large-$k$ limit, the scalar propagation speeds can be obtained as the leading-order solutions of
\be
\det\left(c_s^2{\bm K}-\frac{1}{a^2}{\bm G}
-\frac{i c_s}{a}{\bm B}\right)=0\,,
\label{eq:det_cs}
\ee
which is obtained from a standard WKB scheme with the frequency of the perturbations $\omega^2=c_s^2k^2$ much larger than the time-scale variation of the background quantities. The derivative-mixing matrix ${\bm B}$ contributes at leading order because ${\bm B}=\mathcal{O}(k^0)$ in the large-$k$ limit so its contribution scales as $\omega k\sim k^2$ under the WKB approximation
\cite{Nicolis:2011cs,Aoki:2025bmj}. Since the off-diagonal components of ${\bm K}$ and ${\bm G}$ are suppressed in the large-$k$ limit, the two squared propagation speeds are
\be \label{c2 superflu}
c_{1,2}^2
:=
\frac{1}{2a^2}
\left[
\frac{G_{11}}{K_{11}}+\frac{G_{22}}{K_{22}}
+\frac{B_{12}^2}{K_{11}K_{22}}\pm
\sqrt{
\left(
\frac{G_{11}}{K_{11}}+\frac{G_{22}}{K_{22}}
+\frac{B_{12}^2}{K_{11}K_{22}}
\right)^2
-\frac{4G_{11}G_{22}}{K_{11}K_{22}}
}
\right] \, .
\ee
where $c_1^2$ and $c_2^2$ correspond to the plus and minus branches, respectively.
Using the expressions for ${\bm K}$, ${\bm G}$
and ${\bm B}$ in Appendix \ref{Matrix components}, these become
\ba \label{c1 superflu}
c_{1,2}^2
&=&
\frac{
C_1 bP_b
+\left(bF_b-\frac12\mY F_\mY\right)F_\mX
+\frac14\left(F_\mY+2bs_b\right)^2
}{
2C_1\left(bF_b-\frac12\mY F_\mY\right)
}
\nonumber\\
&&{}\pm
\frac{
\sqrt{
\left[
C_1 bP_b
+\left(bF_b-\frac12\mY F_\mY\right)F_\mX
+\frac14\left(F_\mY+2bs_b\right)^2
\right]^2
-4C_1\left(bF_b-\frac12\mY F_\mY\right)bP_bF_\mX
}
}{
2C_1\left(bF_b-\frac12\mY F_\mY\right)
} \,,
\ea
where $P_b$, $s_b$ and $C_1$ are defined in \cref{thermo_ders,eq:coefs}.
These two propagation speeds have been characterised before
in flat space-time as the first and second sound
\cite{Dubovsky:2011sj,Nicolis:2011cs,Celoria:2017hfd}.
Although the previous expressions may appear relatively uninformative, the following combinations turn out to be considerably simpler
\ba
c_1^2+c_2^2
&=&
\frac{bP_b}{bF_b-\frac12\mY F_\mY}
+\frac{F_\mX}{C_1}
+\frac{(F_\mY+2bs_b)^2}
{4C_1\left(bF_b-\frac12\mY F_\mY\right)}\,,
\\
c_1^2c_2^2
&=& \label{cprod}
\frac{bP_bF_\mX}
{C_1\left(bF_b-\frac12\mY F_\mY\right)}\,.
\ea 
These expressions are nothing but the trace and the determinant of the solutions of \cref{eq:det_cs}.
Note that one of the propagation speeds vanishes if the action is independent of the superfluid component $\mX$, which explains why the invariant $\mY$ alone does not introduce propagating degrees of freedom \cite{Dubovsky:2011sj, Nicolis:2011cs}. This can also be seen from the action \cref{L_flat}; there is no quadratic term containing only spatial
derivatives of $\pi_0$ when $F=F(b,\mY)$.

A general feature is that the fluid and superfluid modes are mixed due to the presence of the derivative-mixing matrix ${\bm B}$.
When $F_\mY+2bs_b=0$, the leading order derivative mixing vanishes \mbox{$B_{12}=B_{21}=0$} and the
speeds reduce to
\be
\bar{c}_1^2=
\frac{bF_b+2b^2F_{bb}}{bF_b-\frac12\mY F_\mY},
\qquad
\bar{c}_2^2=\frac{F_\mX}{C_1}.
\ee
In this case, $\bar{c}_1^2$ depends only on derivatives with respect to the perfect-fluid variables $b$ and $\mY$, while $\bar{c}_2^2$ depends only on derivatives with respect to $\mY$ and $\mX$.
In this sense, the propagation of the perfect-fluid component is described by $\bar{c}_1^2$, while $\bar{c}_2^2$ describes the propagation of the superfluid component.

An interesting case arises when $P_b=F_b+2bF_{bb}=0$, corresponding to a
vanishing leading-order propagation speed for the longitudinal density mode $\bar{c}_1^2=0$. Along with the vanishing derivative matrix constraint \mbox{$F_\mY+2bs_b=0$}, these equations can easily be integrated to give the general solution $F= f_1(\mY) \sqrt{b} + f_2(\mX)$. Notably, the number density-dependent part scales as
$\sqrt{b}$, recovering the usual pressureless perfect-fluid description of cold dark matter when $f_1$ is constant \cite{Celoria:2017xos}. However, more general models with $f_1\neq const.$ or $f_2\neq 0$ can retain this vanishing propagation speed while introducing interactions with the superfluid sector, providing a possible description of interacting dark matter with modified cosmological dynamics. Such models with a non-trivial $\mY$ dependence in $f_1$ were also considered in the recent entropy-coupled works of \cite{Jensko:2026taf}, giving rise to interactions between an exactly-pressureless but entropy-dependent perfect fluid and an external scalar field. 
An entirely different route is to have $bF_b-\frac12 \mY F_\mY\gg bF_b+2b^2F_{bb}$,
leading to a dynamically suppressed propagation speed $\bar{c}_1^2\to 0$. This gives the possibility of obtaining an effective clustering dark matter component through interactions with the superfluid sector, without requiring an exactly pressureless fluid. We will explore this scenario in future work.

Under the no-ghost conditions \eqref{eq:no_ghost}, the two
squared propagation speeds are real and strictly positive
provided
\ba
G_{11}G_{22}&>&0 \,, 
\\
K_{22}G_{11}+K_{11}G_{22}+B_{12}^2&>&0 \,,
\\
\left(K_{22}G_{11}+K_{11}G_{22}+B_{12}^2\right)^2
-4K_{11}K_{22}G_{11}G_{22}&\geq&0 \, .
\ea
Degenerate cases, such as vanishing propagation speeds, require separate consideration \cite{Aoki:2025bmj}.
Together with the no-ghost conditions \eqref{eq:no_ghost}, the following conditions are sufficient to ensure positive quadratic energy and real positive squared propagation speeds \cite{Celoria:2017hfd}
\be \label{eq:gradient_stability}
F_b+2bF_{bb}>0,
\qquad
F_\mX>0.
\ee
These conditions are not strictly necessary for real positive speeds, since the derivative mixing $B_{12}$ could in principle stabilise modes with $G_{11}<0$ or $G_{22}<0$; however, such branches have a quadratic energy that is not bounded from below. In what follows we restrict to the positive-energy case, so that the no-go results below hold for media satisfying \eqref{eq:gradient_stability} in addition to the no-ghost conditions.
The first of these conditions
is nothing but the positivity of the derivative of
the pressure $P_b>0$.
From the second condition and the no-ghost conditions
\cref{eq:no_ghost}, we can derive a generalised null energy
condition for the $F(b,\mY,\mX)$ medium,
\be
\rho+P>2\mY^2F_\mX,
\ee
with the right-hand side strictly positive for
$\mY\neq0$ and $F_\mX>0$.
In the perfect fluid limit, this reduces to the standard
null energy condition $\rho+P>0$.
The absence of instabilities is thus incompatible with phantom-like cosmological behaviour, even in the presence of a superfluid component.
We will explore this result in greater detail in the following section.

Lastly, a key result is obtained from considering the
adiabatic sound speed. The null energy condition together
with \cref{eq:no_ghost,eq:gradient_stability} guarantees
that the adiabatic sound speed is positive,
\be \label{eq:adiabatic_sound_speed}
c_a^2:=\frac{P'}{\rho'}
=
\frac{1}{\rho+P}
\left[
\frac{
\left(F_\mY-2\mY F_\mX-2bF_{b\mY}
+4b\mY F_{b\mX}\right)^2
}{2C_1}
+2bF_b+4b^2F_{bb}
\right]>0.
\ee
This result is somewhat expected since in ordinary perfect
fluids one also obtains $c_a^2>0$ for stability.
The expression for the adiabatic sound speed for the
superfluid medium was obtained in
\cite{Celoria:2017hfd}, while here we have shown that it is
positive under the above stability conditions.
These constraints will be crucial when considering the
slow-roll regime.

\section{No-go results} \label{sec:nogo}
We now investigate the implications of the stability conditions derived above for several cosmologically relevant regimes and concrete models. We first consider the slow-roll regime and show that the stability requirements lead to a no-go result for conventional slow-roll. 
We then focus on phantom crossing and demonstrate that a phantom-to-non-phantom transition for an effective dark energy component is not possible in an expanding universe.
Finally, we apply these criteria to three dark sector models proposed in the literature, providing explicit examples of how the general stability constraints restrict their parameter space.

\subsection{Slow-roll expansion} \label{sec:nogo:slow}

Having understood the dynamics of the propagating degrees
of freedom of the interacting perfect fluid and superfluid in cosmology, we now specialise
to the slow-roll regime and discuss whether the action \eqref{action} can
support slow-roll inflationary solutions while satisfying
the stability conditions.
To study the slow-roll regime we introduce the standard
slow-roll parameters
\be
\epsilon_1
:=
1-\frac{\mathcal H'}{\mathcal H^2}
=
-\frac32
\frac{\mY F_\mY-2\mY^2 F_\mX-2bF_b}
{F-\mY F_\mY-2\mX F_\mX},
\qquad
\epsilon_2:=\frac{\epsilon_1'}{\mathcal H\epsilon_1}.
\label{slow-roll params}
\ee
The background Friedmann and continuity equations imply the exact relation
\be \label{eq:adiabatic_slowroll}
c_a^2
=
-1+\frac23\epsilon_1-\frac13\epsilon_2\, ,
\ee
which can also be seen directly from the expression \eqref{eq:adiabatic_sound_speed}.
Crucially, the stability conditions derived in the previous section guarantee that the adiabatic sound speed is positive, $c_a^2>0$, whereas
 slow-roll gives
\begin{equation}
   \vert \epsilon_1\vert,|\epsilon_2|\ll1 \quad \implies 
c_a^2\simeq-1 \, .
\end{equation}
One can therefore immediately conclude that stable slow-roll is not possible within the general $F(b,\mY,\mX)$
class of models. The relation \eqref{eq:adiabatic_slowroll} was obtained in \cite{Finelli:2018upr} for shift-symmetric cosmologies and in \cite{Aoki:2022ylc} for 2-form cosmologies (which are dual to shift-symmetric scalars), where this obstruction for slow-roll was already pointed out. In this work, we have generalised that result for the general system with a shift-symmetric scalar coupled to a perfect fluid. For a study of the dynamics of inflation of the adiabatic medium $F(b,\mY)$ in different slow-roll regimes, see \cite{Celoria:2019oiu}. This instability can be alleviated when, in addition to the longitudinal mode, transverse modes also propagate, as in solids. In a universe filled by a solid, the propagation speed for longitudinal modes has an extra term coming from the propagating transverse modes: $c_L^2=-1+\frac23 \epsilon_1-\frac13\epsilon_2+\frac43 c_T^2$ \cite{Endlich:2012pz,Gruzinov:2004ty}. Such solid deformations can stabilise the accelerated regime and broaden the range of viable accelerating cosmologies.

The instability can also be seen directly from the
propagation speeds. Using
\cref{cprod},
the product of their squared speeds is
\be \label{eq:sound_product_slowroll}
c_1^2c_2^2
=
\frac{F_\mX}
{2C_1\left(bF_b-\frac12\mY F_\mY\right)}
\left[
(\rho+P)
\left(-1+\frac23\epsilon_1-\frac13\epsilon_2\right)
-
\frac{
\left(F_\mY+2bs_b-2\mY F_\mX\right)^2
}{2C_1}
\right].
\ee
Under the no-ghost conditions and $F_\mX>0$, the prefactor
is positive and $\rho+P>0$.
During slow-roll expansion, the expression in square
brackets is strictly negative, giving $c_1^2c_2^2<0$, implying that one of the squared propagation speeds must be negative and signalling a Laplacian
instability. The product makes this obstruction explicit
without requiring the full expressions for the individual
propagation speeds.

In the perfect fluid limit, the action does not depend
on the superfluid variable $\mX$ and the propagation
speeds are simply
\be \label{c2 slowroll flu}
c_1^2
=\left[ bP_b-\dfrac{(F_\mY-2bF_{b\mY})^2}{2F_{\mY\mY}}\right]
\frac{
1
}{
bF_b-\frac12\mY F_\mY
}
=
-1+\frac23\epsilon_1-\frac13\epsilon_2,
\qquad
c_2^2=0 \, .
\ee
This result is not surprising given that the non-vanishing propagation speed and adiabatic sound speed coincide in the $F(b,\mY)$ limit, therefore automatically ruling out slow-roll in these cases.

Let us stress that the presented obstruction applies to both inflationary models and late-time accelerating dark energy models. Nonetheless, there are some important distinctions to be made: inflation is required to last $\sim 60$ e-folds, so a slow-roll evolution with small $\vert\epsilon_2\vert$ is usually required; on the other hand, late-time dark energy models only require acceleration for $\sim1$ e-fold or less and the requirements on $\epsilon_2$ are less stringent, so the remaining challenge is to construct dark sector models which can satisfy $\epsilon_1<1$ and $\epsilon_2<2\epsilon_1-3$ during the late universe while remaining compatible with cosmological observations. However, the relation \eqref{eq:adiabatic_slowroll} still gives non-trivial restrictions on today's properties of the dark energy sector if we want to avoid instabilities in the transition from matter domination to dark energy domination. Of course, instabilities can arise in the future since we do not have data about that. In summary, our obstruction expressed by \eqref{eq:adiabatic_slowroll} requires to be away from pure slow-roll evolution, but this is not sufficient to guarantee stability. This argument can be made more explicit by studying the background equation of state
\be
w=\frac{P}{\rho}
=-1+\frac23\epsilon_1 \,.
\ee
Accelerated expansion requires $\epsilon_1<1$ while positivity of \cref{eq:adiabatic_slowroll} requires 
$\epsilon_2<2\epsilon_1-3$. From \cref{slow-roll params} we can therefore derive
\be
\frac{\dd\ln H}{\dd\ln a}=-\epsilon_1\,, \qquad
\frac{\dd\ln\epsilon_1}{\dd\ln a}
=\epsilon_2<-1\,,
\ee
where $H=\mathcal H/a$ is the cosmic time Hubble factor.
Thus, $\epsilon_1$ decreases towards zero while the expansion
remains accelerating. If this regular stable evolution
continues indefinitely, $H$ approaches a positive constant
and the background approaches de Sitter.
Stability allows the system to approach the de Sitter limit with $\epsilon_1\to0$ while remaining parametrically far from the standard slow-roll hierarchy $|\epsilon_2|\ll1$. The standard $\Lambda$CDM model provides a concrete example of this, with the late-time limit $\epsilon_2=2\epsilon_1-3\rightarrow -3$, outside of the slow-roll regime but compatible with cosmic observations.
Related behaviour occurs in stable barotropic dark energy
\cite{Linder:2008ya} and in the super slow-roll regime of
fluid inflation \cite{Celoria:2019oiu}.
Exact de Sitter, where $\epsilon_2$ is undefined, must be
considered separately. However, although exact de Sitter scenarios are theoretically interesting, they do not provide a description of realistic cosmological models that always exhibit some departure from it.

We therefore see that any stable approach to a cosmological
constant-like state must be asymptotic, with
$\epsilon_2<-1$, so that $1+w=\frac{2}{3}\epsilon_1$
decays at least as fast as $a^{-1}$. Other than $\Lambda$CDM, a paradigmatic example where precisely this occurs is in the ghost
condensate-type models \cite{Arkani-Hamed:2003pdi,Dubovsky:2005xd},
which possess an exact late-time de Sitter attractor
with a vanishing scalar propagation speed. Finally, it is important to keep in mind the assumptions that lead to \eqref{eq:adiabatic_slowroll}, since the violations of those conditions obviously lead to ways around the discussed obstruction. In this respect, it is worth mentioning that we have  used the propagation speeds computed in the UV limit $k\to\infty$, so models with vanishing propagation speeds at leading order (i.e., at the order we are considering) also need to be treated separately since the next-to-leading order correction to the dispersion relation must be computed. This happens in models that exactly mimic a $\Lambda$CDM background because they identically have $\frac23\epsilon_1-\frac13\epsilon_2=1$ so that $c_a^2=0$. In general, there can be a competition between the next-to-leading order corrections from the large-$k$ expansion and the EFT expansion. Such cases go beyond our present study but represents an interesting future avenue to explore.

\subsection{Phantom crossing}  \label{sec:nogo:phantom}

Next, we consider whether the interacting medium can
cross the phantom divide, $w=-1$, which has long been a topic of interest in extended cosmological models \cite{Kunz:2006wc}. The restrictions on phantom crossing are well-known for simple barotropic fluids \cite{Scherrer:2005je,Linder:2008ya}, and here we extend these results to the general Lagrangian $F(b,\mY,\mX)$. We begin by studying the total equation of state of the medium, before looking at the conditions on \textit{effective} phantom crossing obtained from decomposing the fluid into a cold dark matter and dark energy component. 

The background density and pressure for the total medium satisfy
\be
\rho+P
=
2\left(bF_b-\frac12\mY F_\mY\right)
+2\mY^2F_\mX\, .
\ee
The no-ghost condition \cref{eq:no_ghost},
$bF_b-\frac12\mY F_\mY>0$, together with
$F_\mX>0$ to avoid Laplacian instabilities, therefore gives
\be \label{eq:phantom_bound}
\rho+P>2\mY^2F_\mX>0\,,
\ee
In the perfect fluid limit, this reduces to the standard null energy condition \mbox{$\rho+P>0$}, but in all other cases we find a stricter condition.
For positive energy densities\footnote{Negative energy densities $\rho<0$ allow for a phantom equation of state $w<-1$, but these scenarios are either observationally unviable if $\rho<0$ at all times, or  pathological if accompanied by a transition to positive densities $\rho'>0$, as discussed below.}, this implies 
\be \label{no_phantom}
1+w>\frac{2\mY^2F_\mX}{\rho}>0\,.
\ee
so that the medium must remain above the phantom 
divide $w>-1$. We have therefore shown that including a superfluid component does not allow for phantom crossing without giving rise to pathologies such as ghosts or Laplacian instabilities. Indeed, the superfluid component forces the medium even further away from phantom behaviour than the perfect fluid case. This is nothing but an explicit realisation of the general fact that $N$ scalar fields with up to first order derivatives cannot accommodate a phantom crossing without incurring into instabilities (see e.g. \cite{Dubovsky:2005xd,Rubakov:2014jja}).
  
The superfluid mode contributes positively to the background enthalpy, $\rho+P$, and cannot provide the negative contribution required for phantom crossing. This strengthens the  null energy condition relative to the perfect fluid case $\rho+P>0$. A cosmological constant term in $F$ does not change $\rho+P$ and therefore does not alter this obstruction. However, an asymptotic approach to the phantom divide $w=-1$ is not excluded, since the lower bound on \cref{eq:phantom_bound} can become arbitrarily small.

Finally, the possibility that the energy density of the medium was negative at high redshifts and became positive at low redshifts has received considerable attention as a phenomenological modification of $\Lambda$CDM
\cite{Akarsu:2019hmw,Akarsu:2021fol,Akarsu:2022typ,Soriano:2025gxd,Ibarra-Uriondo:2026zbp, Akarsu:2026pia}. The density parameter for the other cosmological components would therefore have to be greater than one to satisfy the Friedmann constraint.
In this context, it is worth asking whether such a sign change in the energy density can be realised in a superfluid medium while satisfying the stability conditions. For $\rho$ to evolve from a negative to a positive value, there must be a regular transition during which $\rho'>0$. The background continuity equation \eqref{cont_eq} then implies
\be
\rho'=-3\mathcal{H}\left(\rho+P\right)>0\,.
\ee
This inequality is clearly incompatible with the ghost-free condition \eqref{eq:phantom_bound}. We therefore conclude that a sign change in the energy density from negative to positive values inevitably leads to a violation of the stability conditions.

\subsubsection{Effective phantom crossing}
While the above results strictly forbid stable phantom crossing or negative-to-positive sign-switching phenomena, it is straightforward to see that such restrictions are not imposed on an \textit{effective} dark energy component within the $F(b,\mY,\mX)$ superfluid. By decomposing the total medium into a part that evolves like normal cold dark matter $\rho_{\rm CDM} \propto a^{-3}$ and a separately evolving component, we can identify the additional component as an effective dynamical dark energy contribution. 
Such effective phantom behaviour has become a popular exploration in dark sector models in light of the recent DESI results \cite{DESI:2024mwx,DESI:2025zgx}, with numerous potential explanations proposed in the recent literature \cite{Mishra:2026tzn}. Here we show that such a crossing is indeed possible, but the direction is constrained to be from non-phantom to phantom. 

At the background level, we split the total energy density into a pressureless CDM component and an effective dark energy component,
\begin{equation}
    \rho_{\rm CDM} := m \sqrt{b} = \rho_{\rm CDM0} a^{-3}\, , \quad P_{\rm CDM}:=0 \,, \qquad \rho_{\rm DE} := \rho - \rho_{\rm CDM}\, , \quad P_{\rm DE} := P \, ,
\end{equation}
where $m>0$ and $\rho_{\rm CDM0}>0$ are normalisation constants.
This leads to an effective dark energy equation of state
\begin{equation}
    w_{\rm DE} := \frac{P_{\rm DE}} {\rho_{\rm DE}} = \frac{P}{\rho-\rho_{\rm CDM}}\, .
\end{equation}
Since the two components do not interact at the background level, $w_{\rm DE}$ corresponds to the effective dark energy equation of state constrained by the background expansion history in DESI-based cosmological analyses \cite{DESI:2024mwx,DESI:2025zgx}.
Assuming positive densities $\rho_{\rm DE}>0$, $\rho_{\rm CDM}>0$, the condition which prevented stable phantom crossing in the previous section, \cref{no_phantom}, can now be written as
\begin{equation}
    1 + w_{\rm DE}  > \frac{2\mY^2 F_\mX -\rho_{\rm CDM}}{\rho_{\rm DE}} \, ,
\end{equation}
where the right-hand side no longer has to be positive.
Effective phantom crossing is therefore permitted when $\rho_{\rm CDM}>2\mY^2 F_{\mX}$, with the precise details of its realisation depending on the specific physical model. 

Furthermore, because the dark matter and effective dark energy components are individually conserved, 
\begin{equation}
    \rho_{\rm CDM}'=-3\mathcal{H}\rho_{\rm CDM} \,, \quad  \rho_{\rm DE}'=-3\mathcal{H}\rho_{\rm DE} (1+w_{\rm DE}) \,, 
\end{equation}
we can use the definition of the adiabatic sound speed 
$P' = c_a^2 \rho' = - 3 \mathcal{H} c_a^2 (\rho+P)$ from \cref{eq:adiabatic_sound_speed} to obtain
\begin{align}
    w_{\rm DE}' = -3 \mathcal{H} \left[( 1 + w_{\rm DE})(c_a^2 - w_{\rm DE}) + c_a^2 \frac{\rho_{\rm CDM}}{\rho_{\rm DE}} \right] \, .
\end{align}
On the effective phantom divide $w_{\rm DE}=-1$  this reduces to
\begin{equation} \label{phantom_cond}
    w_{\rm DE}'\big|_{w_{\rm DE}=-1} = - 3\mathcal{H} c_a^2 \frac{\rho_{\rm CDM}}{\rho_{\rm DE}} < 0 \,.
\end{equation}
The right-hand side is strictly negative in an expanding universe $\mathcal{H}>0$ because our previous stability results showed that the adiabatic sound speed squared must be positive $c_a^2>0$, and both densities are assumed positive.
This therefore proves that the effective phantom divide can only be crossed from above. Consequently, a stable crossing can only evolve into the phantom regime, and a subsequent return to $w_{\rm DE}>-1$ in an ever expanding universe is excluded while the stability conditions hold.

We therefore obtain a new general restriction on effective phantom crossing in the $F(b,\mY,\mX)$ class of models: no stable effective phantom-to-non-phantom crossing is permitted. In the barotropic limit $F(b)$, these results follow from the established evolution equation for the fluid equation of state \cite{Scherrer:2005je,Linder:2008ya}, but do not appear to have been stated for an \textit{effective} dark energy component. Consequently, all such $F(b,\mY,\mX)$ cosmological media and their limiting cases cannot realise the effective phantom-to-non-phantom transition discussed in connection with recent DESI results \cite{Mishra:2026tzn} while remaining stable.

\subsection{Dark sector models}  \label{sec:nogo:model}

Utilising the propagation speeds and stability conditions of the section above, we can now apply these results to concrete models in cosmology. For this purpose, we investigate three unified dark sector models proposed by Celoria et al. \cite{Celoria:2017xos} that belong to the perfect fluid class described by a Lagrangian of the form $F(b,\mY)$. Furthermore, these models have a vanishing propagation speed associated to the superfluid sector and an adiabatic propagation speed $c_1^2=c_a^2$ associated to the perfect fluid sector. We find that the model in \textit{Case 2} cannot simultaneously be free from ghosts and Laplacian instabilities, while \textit{Case 1} and \textit{Case 3} models can be free from such pathologies. \textit{Case 1} provides an example of how accelerated expansion can occur without obvious instabilities by violating the assumptions of aforementioned obstruction, while \textit{Case 3} avoids instabilities as long as the accelerated slow-roll phase is avoided.

\subsubsection{Case 1 (DM and DE)}
The Lagrangian used in \cite{Celoria:2017xos}, describing dark matter with a cosmological constant-like term, 
is defined by
\be
F=3\Mpl^2H^2_0\left[\Omega_{\rm DM}\sqrt{b}-\Omega_\Lambda(\sqrt{b}\mY)^2\right]\,,
\ee
where $H_0$ is the Hubble constant and $\Omega_{\rm DM}$ and $\Omega_\Lambda$ are the present-day density parameters of dark matter and dark energy. Note that the energy density \eqref{rhoandP_back} for this fluid coincides with the standard dark sector energy density and pressure in the $\Lambda$CDM model,
\be
\rho=3\Mpl^2H^2_0\left(\Omega_{\rm DM}a^{-3}+ \Omega_\Lambda\right)\,,\qquad P=-3\Mpl^2 H_0^2 \Omega_\Lambda\,,
\ee
where we have used the background solution\footnote{The solution of \cref{eq chi} in \textit{Case 1}, $\chi''=4\mathcal{H}\chi'$, is $\chi'=\chi_0 a^4$ with $\chi_0$ being a constant \cite{Celoria:2017xos}. For simplicity and without loss of generality, we can set $\chi_0=1$.} $\chi'=a^4$ of \cref{eq chi}.

In this case, the conditions for absence of ghosts in the small-scale limit, \cref{eq:no_ghost}, imply that the density parameters must be positive 
\be
\Omega_{\rm DM}>0\,, \qquad \Omega_\Lambda>0\,.
\ee
On the other hand, the fluid propagation speed vanishes
\be
c_a^2=c_1^2=0\,.
\ee
This is an example of how our obstruction can be evaded by breaking one of the assumptions since we obtain a vanishing propagation speed in the large-$k$ regime that we have considered. As discussed above, the vanishing of $c_a^2$ was guaranteed by having an exact $\Lambda$CDM background evolution. Thus, establishing the stability of these models requires to compute the next-to-leading order contribution to the propagation speed. Phenomenologically, the vanishing of the propagation speed might be an indication of an instability in the form of a strong coupling since spatial gradients could be excited without any energy cost. In a cosmological context, the Hubble friction can alleviate the instability to a mild Jeans-type of instability, as it occurs for a dust component. However, the vanishing of the propagation speed could also entail more violent instabilities or provide an additional source of cosmological clustering that would conflict with large scale structures observations. In any case, to properly assess these issues, the next-to-leading order contribution needs to be computed.

\subsubsection{Case 2 (Modified DM and DE)}
The second model proposed in \cite{Celoria:2017xos} describes an extended dark sector with the Lagrangian
\be
F=3\Mpl^2H^2_0\left[\frac{1}{3}(\Omega_{\rm DM}+\Omega_z)^{2/3}\left(\Omega_{\rm DM}\frac{\sqrt{b}}{\mY^2}+\frac{\Omega_z}{(\sqrt{b}\mY)^2}\right)+\Omega_\Lambda\right]\,.
\ee
The new $\Omega_z$ component changes the background evolution of the energy density and pressure
\be
\rho=3\Mpl^2H_0^2\left[\Omega_{\rm DM}(1+r)^{2/3}(1+ra^9)^{1/3}a^{-3}+\Omega_\Lambda\right]\,,\quad
P=-3\Mpl^2H_0^2\left[\Omega_z a^6\left(\frac{1+r}{1+ra^9}\right)^{2/3}+\Omega_\Lambda\right]\,.
\ee
where we have defined the fraction $r:= \Omega_z / \Omega_{\rm DM}$ and used the background solution $\chi'=a\left[\Omega_{\rm DM}\left(1+ra^9\right)\right]^{1/3}$ of \eqref{eq chi}. In particular, the presence of $\Omega_z$ modifies the standard matter-like contribution to the energy density by a factor  $(1+r)^{2/3}(1+ra^9)^{1/3}$. This contribution scales as $\Omega_{\rm DM}(1+r)^{2/3}a^{-3}$ at early times and asymptotically approaches a constant value at late times. The pressure also acquires an additional contribution from the $\Omega_z$ sector, which vanishes in the high-redshift limit and becomes constant at late times.

Let us now examine the stability conditions for this extended dark sector model. The conditions for absence of ghosts in the small-scale limit, \cref{eq:no_ghost}, take the form
\be
    \frac{3\Mpl^2H_0^2}{2a\chi'^2}\Omega_{\rm DM}\left(\Omega_{\rm DM}+\Omega_z\right)^{2/3}>0\,, \qquad
    -\frac{3 \Mpl^2 H_0^2 a}{\chi'^4} (\Omega_{\rm DM}+\Omega_za^9) \left(\Omega_{\rm DM}+\Omega_z\right)^{2/3}>0\,.
\ee
In addition, the presence of the $\Omega_z$ term gives rise to a non-vanishing propagation speed for the fluid
\be
c_a^2=c_1^2=2a^9\frac{\Omega_z}{\Omega_{\rm DM}+\Omega_z a^9}\,.
\ee
Note that the no-ghost condition imposes
$\Omega_{\rm DM}+\Omega_z a^9<0$, whereas the positive squared propagation speed requires $\Omega_z<0$. 
Moreover, the first no-ghost condition requires $\Omega_{\rm DM}>0$, so $\Omega_{\rm DM}+\Omega_z a^9<0$ can only hold at late times for $\Omega_z<0$, and it is always violated at early times $a\to0$.
Attempting to satisfy the no-ghost conditions thus leads to a
negative squared propagation speed at early times.
Unlike in \textit{Case 1}, this model cannot remain free from ghosts
and Laplacian instabilities throughout its cosmological evolution,
rendering it intrinsically pathological.

\subsubsection{Case 3 (Exotic dark sector)}
Finally, we analyse the \textit{Case 3} model of \cite{Celoria:2017xos}, defined by
\be
F=3\Mpl^2H_0^2\left(\Omega_{\rm DM} \sqrt{b}-w_z\Omega_z\mY^\frac{1+w_z}{w_z}+\Omega_\Lambda\right)\,,
\ee
where $\Omega_z$ is the present-day density parameter of a component with an effective equation of state $w_z$.
The resulting energy density and pressure can be written as
\be
\rho=3\Mpl^2H_0^2\left(\Omega_{\rm DM}a^{-3}+\Omega_z a^{-3(1+w_z)}+\Omega_\Lambda\right)\,,\quad P=-3 \Mpl^2 H_0^2 \left(-w_z \Omega_z a^{-3(1+w_z)} + \Omega_\Lambda\right)\,,
\ee
where we have used that in this background the solution of \eqref{eq chi} is $\chi'=a^{1-3w_z}$.
We see that the additional $\Omega_z$ component behaves as a barotropic fluid with a constant effective equation of state $w_z$, since $P_z=w_z\rho_z$. Its energy density evolves as $\rho_z\propto a^{-3(1+w_z)}$, or equivalently $\rho_z\propto(1+z)^{3(1+w_z)}$. Thus, the $\Omega_z$ component becomes increasingly relevant towards higher redshifts for $w_z>-1$, whereas it decreases towards the past and becomes dominant at late times for $w_z<-1$. In particular, $w_z=0$ reproduces a matter-like component, $w_z=1/3$ corresponds to radiation, and $w_z=-1$ reduces to a cosmological-constant-like contribution, though these limits need to be taken with care.

The no-ghost conditions are modified by the presence of the barotropic $z$-component,
\be
\frac32 H_0^2\Mpl^2a^{-3}\left(\Omega_{\rm DM}+\left(1+w_z\right)\Omega_z a^{-3w_z}\right)>0\,,\qquad \frac32 H_0^2\Mpl^2\frac{\left(1+w_z\right)}{w_z}\Omega_z a^{-3(1-w_z)}>0\,,
\ee
whereas the pressure derivative $P_b$ appearing in the
Laplacian stability condition, \cref{eq:gradient_stability}, vanishes since the only $b$-dependence comes from the pressureless dark matter component. The no-ghost conditions ensure positive squared propagation speed
\be
c_a^2=c_1^2=\frac{w_z\left(1+w_z\right)\Omega_z a^{-3w_z}}{\Omega_{\rm DM} + \left(1+w_z\right)\Omega_z a^{-3w_z}}\,.
\ee

Let us analyse the different implications for $w_z$ from the stability conditions. These conditions lead to three distinct regimes depending on the value of $w_z$. 
For $w_z>0$, the no-ghost condition requires $\Omega_z>0$, in which case both terms contributing to the first no-ghost condition are positive, ensuring stability for all values of the scale factor.
In the intermediate regime, $-1<w_z<0$, the second no-ghost condition requires $\Omega_z<0$. Although the squared propagation speed remains positive as long as both no-ghost conditions are satisfied, the contribution of the $\Omega_z$ component to the first condition is negative and grows in magnitude towards late times, since $a^{-3w_z}$ grows as the universe expands. Consequently, the condition $\Omega_{\rm DM}+(1+w_z)\Omega_z a^{-3w_z}>0$ is eventually violated.
Finally, for $w_z<-1$, the second no-ghost condition requires $\Omega_z>0$. 
However, since $1+w_z<0$, the $\Omega_z$ contribution to the first no-ghost condition is again negative and grows towards late times, leading to an eventual fatal instability. 
Therefore, although the propagation speed remains positive in the ghost-free regime, only the $w_z>0$ case can remain stable throughout the entire cosmological evolution. In particular, a dark energy-like component with $w_z<-1/3$ cannot provide a fully stable realisation of this model, in compliance with our result \eqref{eq:adiabatic_slowroll} obstructing slow-roll solutions.

In summary, these three examples illustrate how the stability conditions can impose non-trivial constraints on the parameter space of the models. In some cases, they restrict the allowed values and signs of the model parameters, while in others they can exclude entire regions of the parameter space or even rule out a given model as a viable stable realisation.

\section{Conclusions}
\label{sec:conclusions}

In this paper we have studied cosmological scenarios with a matter sector composed of an interacting perfect fluid and superfluid. To study this system, we have employed the EFT formalism of relativistic media, where the superfluid sector is described by a shift-symmetric scalar field, thus making contact with cosmological models involving scalar fields. Furthermore, we have discussed the relation between the EFT approach and classical thermodynamical variables to illustrate how the considered system goes beyond an adiabatic regime.
After presenting the cosmological background evolution, we derived the quadratic action for scalar perturbations, allowing us to directly impose the absence of ghosts and Laplacian instabilities. These requirements led to a set of non-trivial stability conditions that generalise existing results in the literature. In particular, we showed that the positive-energy stability conditions imply a positive adiabatic sound speed squared $c_a^2>0$ and a stricter null energy condition.

A key consequence of our results is that slow-roll is incompatible with stability within the general class of models, since the squared adiabatic sound speed becomes negative in this regime.
This obstruction generalises similar observations made for shift-symmetric cosmologies \cite{Finelli:2018upr} and their dual version in terms of 2-forms \cite{Aoki:2022ylc} to our more general interacting system. Although this result was known for perfect fluids and superfluids independently, we have shown that adding a general interaction between them cannot remedy the obstruction to construct accelerated slow-roll cosmologies with these components. Accelerated expansion, however, is not excluded: stability still allows a stable asymptotic approach to de Sitter, which lies parametrically far from the slow-roll regime. Nonetheless, our results still place strong constraints on accelerating cosmological models outside of the slow-roll regime.
An important caveat of our stability analysis is that we have taken the large-$k$ limit, so models with a non-standard $k^2$ dispersion relation at leading order are not subject to our obstruction. A class of models that in turn fall within this category are those mimicking an exact $\Lambda$CDM background. 

We further showed that phantom crossing is prohibited for the total medium while satisfying the stability conditions, while an effective phantom crossing can only occur in one direction. In particular, the transition can only happen from non-phantom to phantom in an expanding universe. This presents an additional obstruction for constructing stable dark sector models that aim to explain the indications of DESI, which prefer a crossing of the phantom divide in the other direction.

Finally, we have applied our general formalism to proposed unified dark sector scenarios within the perfect fluid subcase, which have appeared previously in the literature. We explicitly illustrate how the stability conditions constrain the values and signs of the parameters, and can even rule out theoretically unstable realisations entirely. This motivates the construction of extended dark sector models with a superfluid component, and a detailed study of their perturbative structure.

Future work could therefore focus on identifying stable models able to produce stable accelerating cosmologies within this formalism. For instance, one route is to investigate interacting superfluid models with clustering behaviour induced by the entropy interactions. A complementary direction is to extend this study towards a statistical and microphysical analysis of superfluids, potentially bridging the gap between their statistical description and their applications to effective dark sector models. Finally, theoretically consistent models need to be confronted and tested with observational data, where the stability conditions derived in this work can be used as model priors.

\begingroup
\makeatletter
\def\addcontentsline#1#2#3{}
\acknowledgments
\makeatother
This work emerged from discussions at the summer school ``Les Houches Summer School on the Dark Universe''. This paper is based upon work from COST Action CA21136 Addressing observational tensions in cosmology with systematics and fundamental physics (CosmoVerse) and COST Action CA24101 Testing fundamental physics with seismology (FuSe) supported by COST (European Cooperation in Science and Technology). This work was financially supported by Grants PID2023-149018NB-C44 and PID2024-158938NB-I00 funded by
MCIU/AEI/10.13039/501100011033 and by “ERDF A way of making Europe” and Project SA097P24. M.P.G. acknowledges support from Fondo Social Europeo Plus, Programa Operativo de Castilla y Le\'on and Junta de Castilla y Le\'on, through the Consejer\'ia de Educaci\'on. 
E.J. is supported by the Engineering and Physical Sciences Research Council (EPSRC) [EP/W524335/1, UKRI3030].
\endgroup

\appendix
\section{Mapping to the thermodynamical variables}
\label{Mapping}
Here we complete the EFT to thermodynamic mapping for the $F(b,\mY,\mX)$ medium, following on from the analysis in \cref{sec:mapping}.
Once we have identified the relative velocity \eqref{xi}, the energy density  \eqref{rhoandP}, pressure \eqref{isoP}, number density \eqref{n} and entropy \eqref{s}, we can determine the remaining relevant thermodynamical variables: the chemical potential $\mu$, the temperature $T$ and the relative fluid velocity conjugate $\eta$. To this end, we choose $n$, $T$ and $\xi$ as the independent variables and work in the \textit{thermodynamical representation} with $\rho=\rho(n,T,\xi)$ and $P=P(n,T,\xi)$. We adopt this representation because the correspondence between the thermodynamical variables and the EFT parameters is unambiguous, unlike in the energy representation $\rho(n,s,\xi)$ \cite{Ballesteros:2016kdx}.

In order to have a complete and consistent mapping between thermodynamic variables and EFT operators, we need to make sure that we satisfy the Euler relation
\be
\rho+P=Ts+\mu n \, , 
\label{Euler}
\ee
and the first law of thermodynamics \cite{landau1987fluid, Nicolis:2011cs}
\be
\text{d}\rho=T \text{d}s + \mu\text{d}n + \eta\text{d}\xi \, ,
\label{diffthermo}
\ee
where $\eta$ is the conjugate variable of $\xi$. Using (\ref{xi}) and (\ref{rhoandP}) we can write Eqs.~(\ref{Euler}) and (\ref{diffthermo}) in terms of the EFT variables
\ba
& &
2bF_b-\mY F_\mY+2\mY^2 F_\mX=T\left(2\mY F_\mX-F_\mY \right)+\mu\sqrt{b}\,,\\
& &
F_b\text{d}b+F_\mX\text{d}\mX+4 \mY F_\mX\text{d}\mY-\mY\text{d}F_\mY+2\mY^2\text{d}F_\mX =T\left(2F_\mX\text{d}\mY+2\mY\text{d}F_\mX-\text{d}F_\mY\right)+\frac{\mu}{2\sqrt{b}}\text{d}b\\
& &
+\frac{\eta}{2\sqrt{\mY^2+\mX}}\left(2\mY\text{d}\mY+\text{d}\mX\right). \nonumber
\ea
These equations hold when we make the following identification
\be
T=\mY\,,\qquad\mu=2\sqrt{b}F_b\,\qquad\eta=2\sqrt{\mY^2+\mX}F_\mX\,.
\ee
We therefore obtain a natural interpretation of $\mathcal{Y}$ as the thermodynamic temperature, and $F_b$ and $F_{\mathcal{X}}$ as relating to the chemical potential and the thermodynamical conjugate of $\xi$, respectively.

Lastly, from the conservation of the currents \eqref{density current} and \eqref{entropy current}, we obtain the evolution of the entropy per particle $\sigma\equiv s/n$ in terms of the energy flux \eqref{heatflux} \cite{Celoria:2017bbh}
\be
u^\alpha \nabla_\alpha \sigma=-\frac{1}{\sqrt{b}}\nabla^\alpha\left(\frac{q_\alpha}{\mY}\right)\,.
\ee
If the Lagrangian \eqref{fluLag} does not depend on the superfluid variable $\mX$ the energy flux vanishes and the entropy per particle is conserved along the fluid flow $u^\alpha \nabla_\alpha \sigma=0$. Therefore, a sufficient condition for adiabaticity is that the effective action does not depend on the variable $\mX$. We therefore conclude that the Lagrangian $F(b,\mY)$ corresponds to an adiabatic but non-barotropic fluid, since the energy density and the pressure \eqref{rhoandP} generally depend on two variables.

With this analysis, we establish a clear dictionary between thermodynamical variables of the general fluid and EFT operators. This dictionary is given in Table \ref{Tabla1}. This allows us to make a classification of the fluid according to the symmetries of the Lagrangian.

\begin{table}[h!]
\centering
\renewcommand{\arraystretch}{2}
\begin{tabular}{|c|c|}
\hline
EFT Operators & Thermodynamical variables \\
\hline\hline
$\sqrt{b}$ & $n$ number density \\
\hline
$2\mY F_\mX-F_\mY$ & $s$ entropy density 
\\
\hline
$\mY$ & $T$ temperature \\
\hline
$2\sqrt{b}F_b$ & $\mu$ chemical potential \\
\hline
$\sqrt{\mY^2+\mX}$ & $\xi$ relative velocity \\
\hline
$2\sqrt{\mY^2+\mX}F_\mX$ & $\eta$ conjugate to $\xi$ \\
\hline
$2h^{\mu\alpha}\omega_\alpha\sqrt{-\mX}\mY F_\mX$ & $q^\mu$ energy flux\\
\hline
$\sqrt{b}u^\mu$ & $J_n^{\mu}$ particle number density current \\
\hline
$\left(2\mY F_\mX-F_\mY\right) u^\mu+\frac{q^\mu}{\mY}$ & $J_s^{\mu}$ entropy density current \\
\hline
\end{tabular}
\caption{Summary of the dictionary between the EFT operators and thermodynamical variables, in the representation where $n$,  $T$ and $\xi$ are independent variables.}
\label{Tabla1}
\end{table}

Finally, it is worth commenting on the relation to other well-studied formalisms in the literature. 
In this work, we have employed the EFT framework to describe relativistic fluids, where the thermodynamic variables are interpreted as effective fluid variables. An equivalent formulation for describing relativistic perfect fluids in cosmology is the Lagrangian formalism pioneered by Schutz and Sorkin \cite{Schutz:1977df} and Brown \cite{Brown:1992kc}. This formalism has been applied to non-minimally coupled fluid models \cite{Pourtsidou:2013nha, Skordis:2015yra, Koivisto:2015qua, Boehmer:2015kta, Boehmer:2015sha, Dutta:2017kch, Boehmer:2024rqk, Ashi:2025dba, Boehmer:2026syd} and has recently been extended to include intrinsic entropic perturbations \cite{Jensko:2026taf}.
The particle-number conservation equation and the entropy conservation equation are obtained by varying the action with respect to the corresponding Lagrange multipliers. Consequently, both currents are conserved on-shell. This contrasts with the EFT approach, where one of the conservation equations is satisfied off-shell due to the number of degrees of freedom and symmetries being fixed by construction.

\section{Vector and tensor perturbations}
\label{Append1}
In this Appendix, we include the perturbations for the vector and tensor sector.
For the gravitational sector, we consider the perturbed line element \cite{MUKHANOV1992203}
\be
{\rm d}s^2 = -a^2(1 + 2A){\rm d}\eta^2 
+ 2a^2 \left(2\partial_{i}B+S_i\right){\rm d}\eta {\rm d}x^i
+ a^2 \Big[ (1 - 2\zeta) \delta_{ij}
+ 2 \left(\partial_i \partial_j -\frac13 \delta_{ij}\nabla^2\right) E +\partial_i F_j+\partial_j F_i + h_{ij}\Big]
{\rm d}x^i {\rm d}x^j \,,
\ee
where $A$, $B$, $\zeta$ and $E$ describe scalar perturbations, $S_i$ and $F_i$ are transverse vectors which satisfy \mbox{$\partial_i S^i=\partial_i F^i=0$}, and $h_{ij}$ is a transverse-traceless 
tensor perturbation with
\mbox{$\partial^i h_{ij}=h^i{}_i=0$}.

We will work in Fourier space writing a generic perturbation $\mathcal{A}$ in real space according to
\begin{equation}
\mathcal{A}(\eta,\mathbf{x})
=
\int\frac{\dd^3k}{(2\pi)^3}\,
\mathcal{A}_{\mathbf{k}}(\eta)e^{i\mathbf{k}\cdot\mathbf{x}} \,,
\end{equation}
with the reality condition $\mathcal{A}_{-\mathbf{k}}=\mathcal{A}_{\mathbf{k}}^{*}$.
\subsubsection{Tensor modes}
Expanding the EFT scalars \eqref{scalars} for tensor perturbations up to second order we obtain
\be
b=\frac{1}{a^6}\left(1+\frac12 h_{ij}h^{ij}\right)\,,\qquad \mY=\frac{\chi'}{a}\,,\qquad\mX=-\frac{\chi'^2}{a^2}\,.
\ee
Since the perturbations of the internal coordinates do not contain tensor modes, tensor perturbations of the metric enter exclusively through the determinant $b$. Substituting these expressions into the action \eqref{action}, expanding up to second order, performing integrations by parts and imposing the background equations of motion Eqs.~(\ref{eq chi}), (\ref{eqsback}) and (\ref{eqsback2}), we obtain the quadratic action for tensor perturbations
\begin{equation}
{\cal S}^{(2)}_{\text{tensor}}=  \frac{1}{64\pi G} \int \frac{{\rm d}\eta {\rm d}^3 k}{(2\pi)^3}\, a^2
\left( h_{ij}'^2 - k^2 h_{ij}^2 \right) \,.
\label{St}
\end{equation}
The propagation of gravitational waves is governed by the usual action in FLRW background, with a propagation speed equal to the speed of light. Ghost and Laplacian instabilities are absent as a direct consequence of the Einstein-Hilbert term and the lack of non-minimal couplings. Note that the latter expression is not modified by the addition of other perfect fluids (the matter sector for example), since they do not contribute to the tensor sector at linear order.

We have just shown that the dynamics of gravitational waves in superfluids remains identical to that in standard General Relativity. This is not the case for solids, where the shear response gives rise to a time-dependent effective mass for GWs and a non trivial gravitational slip \cite{Celoria:2017bbh, BeltranJimenez:2026tlr}. We now move on to the vector sector, which is sensitive to the superfluid components.

\subsubsection{Vector modes}
We will focus now on the vector perturbations in the gauge in which the contribution of $\partial_iF_j{\rm d}x^i {\rm d}x^j$ vanishes. The vector modes of \cref{scalars} are
\be
b=\frac{1}{a^6}+\frac{1}{a^6}\left(-S_i^2+2S_i\delta^i_a\pi_T^{a\prime}-|\vec{\pi}_T'|^2\right)\,,
\quad
\mY=\frac{\chi'}{a}+\frac{\chi'}{2a}\left(|\vec{\pi}_T'|^2-2S_i\delta^i_a\pi_T^{a\prime}\right)\,,
\quad
\mX=-\frac{\chi'^2}{a^2}+\frac{\chi'^2}{a^2}S_i^2\,,
\ee
with $\vec{\pi}_T=\pi_T^a$.
After integrating out the non-dynamical perturbation $S_i$, performing integrations by parts and using the background equations, the quadratic action for 
the transverse mode reads
\be
{\cal S}^{(2)}_{\text{vector}}=\int \frac{{\rm d}\eta {\rm d}^3 k}{(2\pi)^3}
a^4\frac{bF_b-\frac12\mY F_\mY}{1+16\pi Ga^2/k^2\left(2bF_b-\mY F_\mY\right)}
\vert\vec{\pi}_T'\vert^2\,.
\label{Sv}
\ee
We find that the transverse modes do not propagate, as in the perfect fluid case. The ghost-free condition \cref{eq:no_ghost} obtained from the scalar perturbations guarantees the positivity of the kinetic term for the transverse mode.

\section{Matrix components}
\label{Matrix components}
Here we provide the expressions for the components of the matrices appearing in the action \cref{eq:quadratic_action}. We fix the Spatially Flat Slicing Gauge (SFSG)
\be
{\rm d}s^2 = -a^2(1 + 2A){\rm d}\eta^2 
+ 4a^2 \partial_{i}B{\rm d}\eta {\rm d}x^i
+ a^2 \delta_{ij}
{\rm d}x^i {\rm d}x^j \,.
\label{pertmetric_SFSG}
\ee

At leading order in the large-$k$ expansion, the matrix components are given by
\ba
& &
K_{11}=bF_b-\frac12\mY F_\mY+{\cal O}(k^{-1})\,,\quad 
K_{22}=a^{-2} \left(F_\mX -\frac12 F_{\mY\mY} + 2\mY F_{\mY\mX} - 2\mY^2F_{\mX\mX} \right)+{\cal O}(k^{-1})\,, \\
& &
K_{12}=K_{21}=-\frac{\chi'}{\Mpl^2\mathcal{H}k}\left(bF_b-\frac12 \mY F_\mY\right)\left(F_\mX -\frac12 F_{\mY\mY} + 2\mY F_{\mY\mX} - 2\mY^2F_{\mX\mX} \right)+{\cal O}(k^{-2})\,, \\
& &
G_{11}=a^{-4} \left(F_b +2bF_{bb}\right)+{\cal O}(k^{-1})\,,
\quad 
G_{22}=F_\mX+{\cal O}(k^{-1})\,,
 \\
& &
G_{12}=G_{21}
=\frac{1}{12\Mpl^2\mathcal{H}k}
\Bigg\{
4a^{-2}\chi'^{-1}FF_b
-4a^{-3}(F_b F_\mY +FF_{b\mY})
+9a\beta \Mpl ^ 2 \mathcal{H} ^ 2 \mY \left(2\mY F_{\mY\mX} - F_{\mY\mY}\right)
\nonumber
\\
& &
+72a\beta \Mpl^2\mathcal{H}^2 b\mY^2 \left(F_{b\mY\mX}-\mY F_{b\mX\mX}\right)
-4a^{-4}\left[
3\Mpl^2\mathcal{H}^2\chi'{}^{-1}F_b
+a\mY(F_b F_\mX-2 FF_{b \mX}-F_\mY F_{b \mY})
\right]
\nonumber
\\
& &
-4a^{-5}\left[
6\Mpl^2\mathcal{H}^2F_{b\mY}
+\chi'{}^2\left(2F_{b\mX}F_\mY- F_\mX F_{b\mY}\right)
\right]
+2ab\mY\Big[
-4a^2\mY^2 F_\mX F_{b\mX}
+3\Mpl^2\mathcal{H}^2
\big((2+6\beta)F_{b\mX}-3\beta F_{b\mY\mY}\big)
\Big]
\nonumber
\\
& &
+9a\Mpl^2\mathcal{H}^2\left[F_\mY+4b^2\left(2\mY F_{bb\mX}
-F_{bb\mY}\right)\right]
\Bigg\}+{\cal O}(k^{-2}) \,,
\\
& &
B_{12}=-B_{21}=\frac{1}{2}\left(F_\mY-2bF_{b\mY}+4b\mY F_{b\mX}\right)+{\cal O}(k^{-1})\,.
\ea
Note that the $\beta$ term in $G_{12}$ and $G_{21}$ comes from imposing the background equation of motion \cref{eq chi}.

\bibliography{bib}

\end{document}